\documentclass[letterpaper, 10 pt, conference]{ieeeconf}  % Comment this line out if you need a4paper

\IEEEoverridecommandlockouts                              % This command is only needed if 
\usepackage{acronym}
\usepackage{xparse} % Required for \NewDocumentCommand
\usepackage{amsmath,amssymb,amsfonts}
\usepackage{algorithmic}
\usepackage{graphicx}
\usepackage{textcomp}
\usepackage{color}
\usepackage{xspace}
\usepackage{acronym}
\usepackage{graphicx}
\usepackage{lettrine}
\usepackage{amssymb}% http://ctan.org/pkg/amssymb
\usepackage{pifont}%
\usepackage{multirow}
\usepackage{array}

\usepackage{rotating}
\usepackage{bigstrut}
\usepackage{colortbl}
\usepackage{booktabs}
\usepackage{comment}
\usepackage{stackengine}

\usepackage[ruled,linesnumbered,algo2e] {algorithm2e}
\usepackage{hyperref}

\SetKwFor{Loop}{loop}{}{end loop}

\title{\LARGE \bf
ecg2o: A Seamless Extension of g2o for \\ Equality-Constrained Factor Graph Optimization*
}

\author{Anas Abdelkarim$^{1}$, Holger Voos$^{2}$, and Daniel Görges$^{3}$% <-this % stops a space
\thanks{*This research was funded in whole, or in part, by the Luxembourg National Research 
Fund (FNR), MOCCA Project, ref. 17041397. For the purpose of open access, and in fulfilment of the obligations arising from the grant agreement, the author has applied a Creative Commons Attribution 4.0 International (CC BY 4.0) license to any  Author Accepted Manuscript version arising from this submission.}% <-this % stops a space
\thanks{$^{1}$Anas Abdelkarim is with the Interdisciplinary Center for Security, Reliability and Trust (SnT), University of Luxembourg, L-1855 Luxembourg, Luxembourg, and the Department of Electrical and Computer Engineering, RPTU University Kaiserslautern-Landau, 67663, Germany.  
{\tt\small anas.abdelkarim@uni.lu, abdelkarim@eit.uni-kl.de}}%
\thanks{$^{2}$Holger Voos is with the Interdisciplinary Center for Security, Reliability and Trust (SnT), University of Luxembourg, L-1855 Luxembourg, Luxembourg, and the Faculty of Science, Technology, and Medicine (FSTM), Department of Engineering, L-1359 Luxembourg, Luxembourg.  
        {\tt\small holger.voos@uni.lu}}%
\thanks{$^{3}$Daniel Görges is with the Department of Electrical and Computer Engineering, RPTU University Kaiserslautern-Landau, 67663, Germany.  
        {\tt\small goerges@eit.uni-kl.de}}%
}

\begin{document}

\maketitle
\thispagestyle{empty}
\pagestyle{empty}

%%%%%%%%%%%%%%%%%%%%%%%%%%%%%%%%%%%%%%%%%%%%%%%%%%%%%%%%%%%%%%%%%%%%%%%%%%%%%%%%
%% Acronyms
% mathematics symbols
%\acrodef{name}[description]{$value \ensuremath{\mathcal{X}}} % to call use   \ac{name} = value (description) ... \acl{name} = value ..\acs{name} = description
% \acrodef{mathX}[ff]{\ensuremath{\mathcal{X}}}
\newcommand{\matheq}[1]{ {$#1$}}
\newcommand{\mathX}[0]{\ensuremath{X}}
\newcommand{\mathXc}[0]{\ensuremath{Xc}}
\newcommand{\mathU}[0]{\ensuremath{U}}
\newcommand{\mathx}[0]{\ensuremath{x}}
\newcommand{\mathu}[0]{\ensuremath{u}}
\newcommand{\mathQ}[0]{\ensuremath{\text{Q}}}
\newcommand{\mathR}[0]{\ensuremath{\text{R}}}
\newcommand{\mathstep}[1]{\ensuremath{\Delta{\mathX}_{#1}}}
%\NewDocumentCommand{\step}{O{\mathX} O{}}{\ensuremath{\Delta{#1}_{#2}}}
\NewDocumentCommand{\step}{O{\mathX} o}{%
    \IfValueTF{#2}{%
        \ensuremath{\Delta{#1}_{\text{#2}}}%
    }{%
        \ensuremath{\Delta{\mathX}_{\text{#1}}}%
    }%
}

\NewDocumentCommand{\mathJacobian}{m g}{%
    \IfNoValueTF{#2}{%
      \ensuremath{\mathcal{J}_{#1}}
    }{%
      \ensuremath{\mathcal{J}_{#1}(#2)}
    }%
}

\NewDocumentCommand{\mathHessian}{m g}{%
    \IfNoValueTF{#2}{%
      \ensuremath{\text{H}_{#1}}
    }{%
      \ensuremath{\text{H}_{#1}(#2)}
    }%
}
\newcommand{\mathXMAP}[0]{\ensuremath{X^{\text{MAP}}}}
\newcommand{\mathdiag}[1]{\ensuremath{\textbf{diag}(#1)}}
\newcommand{\mathXinitial}[1]{\ensuremath{\widetilde{#1}}}
\newcommand{\mathE}[0]{\ensuremath{\mathcal{E}}}
\newcommand{\mathF}[0]{\ensuremath{\mathcal{F}}}
\newcommand{\mathfactor}[0]{\ensuremath{\phi}}
\newcommand{\mathmatrix}[2]{\ensuremath{\left\lbrack \begin{array}{#1} #2 \end{array} \right \rbrack}}

\newcommand{\mathmean}[0]{\ensuremath{\boldsymbol{\mu}}}
\newcommand{\mathbrk}[1]{\ensuremath{{\bigg[ #1 \bigg] }}} % bracket
\newcommand{\mathprn}[1]{\ensuremath{\left(#1\right)}} % parenthesis
\NewDocumentCommand{\func}{mo}{\ensuremath{#1\IfValueT{#2}{\mathprn{#2}}}}  % \func{f}[x] = f(x)
\NewDocumentCommand{\norm}{m O{2}}{\ensuremath{\left\lVert#1\right\rVert}^2_{#2}}
\newcommand{\mathnormdist}[0]{\ensuremath{\mathcal{N}}} %normal distribution
\newcommand{\mathinormdist}[0]{\ensuremath{\sim\mathcal{N}}} %normal distribution}
\newcommand{\mathedge}[0]{\ensuremath{{\varepsilon}}}
\newcommand{\matherr}[0]{\ensuremath{{e}}}
\newcommand{\mathcov}[0]{\ensuremath{{\Sigma}}}
\newcommand{\mathinfo}[0]{\ensuremath{{\Omega}}}
\newcommand{\mathtrs}[1]{\ensuremath{{#1}^\text{T}}}% transpose
\newcommand{\mathargmax}[1]{\ensuremath{{\underset{#1}{\text{argmax}}}}}
\newcommand{\mathargmin}[1]{\ensuremath{{\underset{#1}{\text{argmin}}}}}
\newcommand{\mathminimize}[1]{\ensuremath{\underset{#1}{\operatorname{min}}}}
\newcommand{\mathmaximize}[1]{\ensuremath{\underset{#1}{\operatorname{max}}}}
\newcommand{\mathindexset}[1]{\ensuremath{\mathcal{I}_{#1} }}

\newcommand{\txtand}[0]{\ensuremath{\text{ and }}}
\newcommand{\txtor}[0]{\ensuremath{\text{ or }}}
 \newcommand{\deltemp}{\let\temp\undefined} % delete temp

\newcommand{\neqc}{\ensuremath{\text{l}}}
 \newcommand{\nineq}{\ensuremath{\text{q}}}
\newcommand{\nJc}{\ensuremath{\text{p}}}
\newcommand{\nXc}{\ensuremath{\text{N}}}
\newcommand{\Leqc}{\ensuremath{\nu}}
\newcommand{\Lineq}{\ensuremath{\lambda}}
\newcommand{\slack}{\ensuremath{s}}

\acrodef{NN}{Neural Network}
\acrodef{RoI}{Region of Interest}
\acrodef{RTK}{Real-Time Kinematic}
\acrodef{SA}{Situational Awareness}
\acrodef{EKF}{Extended Kalman Filter}
\acrodef{DSG}{Dynamic Scene Graph}
\acrodef{UKF}{Unscented Kalman Filter}
\acrodef{VIO}{Visual-Inertial Odometry}
\acrodef{MCL}{Monte Carlo Localization}
\acrodef{MHE}{Moving Horizon Estimation}
\acrodef{GPS}{Global Positioning System}
\acrodef{SURF}{Speeded-Up Robust Features}
\acrodef{CNN}{Convolutional Neural Network}
\acrodef{RNN}{Recurrent Neural Network}
\acrodef{GNN}{Graph Neural Network}
\acrodef{LIDAR}{Light Detection and Ranging}
\acrodef{ATIS}{Asynchronous Time-based Image Sensor}
\acrodef{DAVIS}{Dynamic and Active-pixel Vision Sensor}
\acrodef{AV}{Autonomous Vehicle}
\acrodef{SDF}{Signed Distance Field}
\acrodef{TSDF}{Truncated Signed Distance Field}
\acrodef{ESDF}{Euclidean Signed Distance Field}

\newcommand{\orcid}[1]{\href{https://orcid.org/#1}{\textsuperscript{\includegraphics[height=3mm]{Figures/orcid.pdf}}}}
\newcommand{\orcidAnas}[0]{\orcid{0000-0003-4947-8809}}
\newcommand{\orcidHolger}[0]{\orcid{0000-0002-9600-8386}}
\newcommand{\orcidDaniel}[0]{\orcid{0000-0001-5504-0972}}

\newcommand{\wrt}{w.r.t. }
\newcommand{\cmark}{\ding{51}}
\newcommand{\cxmark}{\ding{51}}
\newcommand{\xmark}{\ding{55}}
\newcommand{\cross}{\ding{55}}
\newcommand{\eg}{\textit{e.g., }}
\newcommand{\ie}{\textit{i.e., }}
\newcommand{\etal}{\textit{et al. }}
\newcommand{\newtxt}[1]{\textcolor{purple}{{\em #1}}}
\newcommand{\ali}[1]{\textcolor{orange}{{\bf [Ali: }{\em #1}{\bf ]}}}
\newcommand{\hriday}[1]{\textcolor{blue}{{\bf [Hriday: }{\em #1}{\bf ]}}}
\newcommand{\claudio}[1]{\textcolor{violet}{{\bf [Claudio: }{\em #1}{\bf ]}}}
\newcommand\claudiost{\bgroup\markoverwith{\textcolor{violet}{\rule[0.5ex]{2pt}{0.4pt}}}\ULon}
\newcommand\alist{\bgroup\markoverwith{\textcolor{orange}{\rule[0.5ex]{2pt}{0.4pt}}}\ULon}
\newcommand\hridayst{\bgroup\markoverwith{\textcolor{blue}{\rule[0.5ex]{2pt}{0.4pt}}}\ULon}

%%%%%%%%%%%%%%%%%%%%%%%%%%%%%%%%%%%%%%%%%%%%%%%% 
% for tables  colomn length
% p{#1}: Multiline column with fixed width and top alignment
%b{#1}: Multiline column with fixed width and bottom alignment
%P{#1}: Custom paragraph column with full justification
%Q{#1}: Custom paragraph column with right alignment
%m{#1}: Multiline column with vertical centering.
%\hspace{0pt}: This allows hyphenation within the cell.
% >{\raggedright\arraybackslash}p{0.13\linewidth}
%\newcolumntype{L}[1]{>{\raggedright\let\newline\\\arraybackslash\hspace{0pt}}m{#1\linewidth}}
%\nohyphenation  % \let\newline\\

\newcolumntype{L}[1]{>{\raggedright\arraybackslash\hspace{0pt}}p{#1\linewidth}}
\newcolumntype{M}[1]{>{\centering\arraybackslash\hspace{0pt}}p{#1\linewidth}}
\newcolumntype{R}[1]{>{\raggedleft\arraybackslash\hspace{0pt}}p{#1\linewidth}}
%\newcolumntype{C}[1]{>{\centering\arraybackslash}p{#1}}
%\newcolumntype{L}[1]{>{\raggedright\arraybackslash}p{#1}}
%% the w=width of the row
\newcommand\xrowht[2][0]{\addstackgap[.5\dimexpr#2\relax]{\vphantom{#1}}}
%\hline\xrowht[()]{10pt}
\NewDocumentCommand{\defaulttable}{ O{1.5} O{2pt}  O{10pt} O{.5pt} O{0pt} O{0pt}    }{
    \renewcommand{\arraystretch}{#1} % the stretch of the table vertically
    \setlength{\tabcolsep}{#2} % the distance from the beginning of the cell
    \setlength{\extrarowheight}{#3} % Extra height vertical above in the rows
    \setlength{\aboverulesep}{#5} % the empty space above the rule for booktabs library
    \setlength{\belowrulesep}{#6} % the empty space below the rule    for booktabs library
     }
%%%%%%%%%%%%%%%%%%%%%%%%%%%%%%%%%%%%%%%%%%%%%%%%

% new environment 
\newenvironment{editor}
{
    %\clearpage % Start on a new page
    
   % \noindent % Optional: prevent indentation at the start
    
}
{
   %  \clearpage % End the environment with a new page   
}

\newcommand{\highlight}[1]{
%\textcolor{red}{#1}
#1
}

\begin{abstract}
Factor graph optimization serves as a fundamental framework for robotic perception, enabling applications such as pose estimation, simultaneous localization and mapping (SLAM), structure-from-motion (SfM), and situational awareness. Traditionally, these methods solve unconstrained least squares problems using algorithms such as Gauss-Newton and Levenberg-Marquardt. However, extending factor graphs with native support for equality constraints can improve solution accuracy and broaden their applicability, particularly in optimal control. In this paper, we propose a novel extension of factor graphs that seamlessly incorporates equality constraints without requiring additional optimization algorithms. Our approach maintains the efficiency and flexibility of existing second-order optimization techniques while ensuring constraint feasibility. To validate our method, we apply it to an optimal control problem for velocity tracking in autonomous vehicles and benchmark our results against state-of-the-art constraint handling techniques. Additionally, we introduce ecg2o, a header-only C++ library that extends the widely used g2o factor graph library by adding full support for equality-constrained optimization. This library, along with demonstrative examples and the optimal control problem, is available as open source at
\href{https://github.com/snt-arg/ecg2o}{https://github.com/snt-arg/ecg2o}.

\end{abstract}

%%%%%%%%%%%%%%%%%%%%%%%%%%%%%%%%%%%%%%%%%%%%%%%%%%%%%%%%%%%%%%%%%%%%%%%%%%%%%%%%
\section{Introduction}
Factor graphs are extensively used in robotic perception tasks for efficiently modeling large-scale probabilistic inference problems \cite{dellaert2017factor}. These methods use graph-based optimization techniques, such as weighted least squares, to address key problems like SLAM and situational awareness \cite{tourani2022visual, bavle2022s}. Typically, these optimization problems are addressed using second-order unconstrained algorithms, including Gauss-Newton  \cite{lai2017solving} and Levenberg–Marquardt \cite{more2006levenberg}. Several well-established libraries, such as GTSAM \cite{gtsam_}, g2o \cite{kummerle2011g}, and SRRG2 \cite{grisetti2020least}, have been developed to provide robust back-end implementations of these optimization algorithms, along with user-friendly front-end interfaces tailored for robotic applications.

Recent advancements in the literature have extended factor graph–based optimization beyond perception tasks to encompass optimal control applications. The primary challenge in this extension lies in handling constraints, as optimal control problems inherently involve constraints, unlike perception problems, which are typically unconstrained. However, using factor graphs for control has a significant advantage: it enables seamless integration between perception and control in robotics, creating a unified optimization framework. Additionally, this approach facilitates the reuse of established algorithms and computational efficiencies available in factor graph libraries \cite{abdelkarim2025factor}. In contrast, traditional optimization frameworks widely used in optimal control, such as CasADi \cite{andersson2019casadi}, AMPL \cite{abdelkarim2020optimal, abdelkarim2023optimization}, and IPOPT \cite{wachter2006implementation}, although highly effective, do not naturally integrate with factor graph–based estimation frameworks. This disconnect can introduce computational inefficiencies and added complexity, particularly in robotic applications that demand a seamless integration of perception and control.

It is important to highlight that, in optimal control problems, equality constraints define deterministic relationships between variables, such as system dynamics that strictly govern state transitions. In contrast, in SLAM, the motion and sensor models establish probabilistic relationships, where uncertainty is inherently modeled and optimized within the factor graph framework. Therefore, equality constraints in factor graphs enforce exact dependencies between variables, ensuring that the solution adheres to physically consistent state transitions, whereas probabilistic constraints model uncertainties and allow for flexible estimation

\subsection{Related Work}
In our recent publication, we presented a comprehensive review of factor graphs in optimal control problems, along with various methods for handling constraints \cite{abdelkarim2025factor}. Here, we provide an overview of the key literature in this area.

Chen and Zhang \cite{chen2019LQR} introduced a general framework for applying Linear Quadratic Regulators (LQR) using factor graphs. In this framework, equality constraints are used to enforce system dynamics and are incorporated into the cost function as a weighted least squares term. By assigning a sufficiently large weighting matrix, these equality constraints effectively act as soft constraints, which are approximately satisfied as the weighting approaches infinity.

Yang et al. \cite{yang2021equality} extended this approach by introducing additional equality constraints between variables within the LQR framework.
Factor graph–based LQR has been applied to a variety of domains, including wireless mesh network control \cite{darnley2021flow}, tactile estimation and extrinsic contact control \cite{kim2023simultaneous}, and trajectory generation for graffiti robots \cite{chen2022gtgraffiti, chen2022locally}. Notably, these implementations have primarily relied on the GTSAM library to formulate and solve the optimization problems.

Beyond LQR-based approaches, more advanced constrained optimization techniques have been explored for handling equality constraints, most notably the Augmented Lagrangian (AL) method \cite{bertsekas2014constrained}. The fundamental idea of AL is to introduce Lagrange multiplier terms and a quadratic penalty term for constraint violations into the cost function, thereby forming the Augmented Lagrangian function. The optimization process involves two nested loops: the inner loop minimizes the Augmented Lagrangian function while keeping the Lagrange multipliers fixed, while the outer loop updates the Lagrange multipliers and penalty parameters until a specified stopping criterion is met.

The AL method has been implemented in GTSAM for improved state estimation \cite{sodhi2020ics} and has been applied in navigation and 3D manipulation planning \cite{qadri2022incopt}. Furthermore, AL has been utilized in SRRG2 for localization and control of unicycle robots \cite{bazzana2022handling} and for pseudo-omnidirectional platform control \cite{bazzana2024augmented}.

While these methods have been successfully applied to various robotics and control problems, existing approaches still suffer from key limitations as described below.

\subsection{Motivation}
Incorporating equality constraints into factor graphs has been primarily achieved through two approaches: soft constraints and the Augmented Lagrangian (AL) method. In the soft constraints approach, enforcing equality constraints requires assigning an infinitely large weighting matrix, which is not feasible in practical implementations. Instead, a large but finite weighting matrix is typically used. However, choosing an appropriate weight requires careful tuning to strike a balance between effectively enforcing the constraint and avoiding ill-conditioned optimization problems. If the original optimization problem is already ill-conditioned, determining an appropriate weighting matrix becomes even more challenging.

In contrast, the Augmented Lagrangian (AL) method systematically enforces equality constraints without requiring direct weight tuning. However, AL has notable limitations. First, as previously mentioned, it employs a nested loop structure, which may lead to unnecessary iterations in the inner loop. Second, the performance of the AL method is highly dependent on hyperparameter tuning, including the initial penalty term, penalty update factor, maximum penalty value, inner-loop iteration limit, and stopping criteria. Suboptimal parameter selection can significantly degrade performance.

To address these limitations, we propose an alternative approach that extends the Gauss-Newton and Levenberg–Marquardt methods, which are the primary unconstrained optimization algorithms used in factor graphs, by incorporating Karush-Kuhn-Tucker (KKT) conditions to explicitly enforce equality constraints. This approach offers faster convergence and eliminates the need for extensive hyperparameter tuning. Furthermore, despite g2o’s efficiency in robotic perception tasks, it does not provide built-in mechanisms for explicitly handling equality constraints. This limitation necessitates either manual constraint encoding using soft constraints (which leads to tuning challenges) or extending the solver framework itself.

\subsection{Contributions}
Our contributions are summarized as follows:

 \begin{enumerate}
    \item Algorithmic Contribution
     \begin{itemize}
     \item We propose a KKT system-based Gauss-Newton method for efficiently handling equality constraints within factor graphs.
    \item We formulate the implementation of equality constraints using the proposed method by representing them as factor graph edges.
   \end{itemize}

    \item Implementation \begin{itemize}
         \item We develop a header-only C++ library (\texttt{ecg2o}) that extends the g2o optimization framework. This library supports both the Augmented Lagrangian method and the proposed KKT-based method. It is publicly available at \url{https://github.com/snt-arg/ecg2o} for reproducibility and further research.       
          %  , enabling seamless switching between algorithms. \item We provide an intuitive interface for defining equality constraint edges, making their implementation as straightforward as standard factor graph edges
         \item We implement a stopping criterion based on the norm of the update step within g2o, ensuring that iterations terminate when step sizes become sufficiently small.
    \end{itemize}

   \item Validation: We provide a case study demonstrating the effectiveness of our method in trajectory tracking optimal control, validating the approach in a real-world robotics application.
 \end{enumerate}

\section{Preliminaries}
\label{sec:preliminary}
Factor graphs consist of variable nodes and factor nodes (also called edges) and represent a factorization of probability density functions (PDFs), which are typically assumed to follow a Gaussian distribution. In maximum a posteriori (MAP) inference, the objective is to maximize the factorized probability function, mathematically formulated as
\newcommand{\temp}{\func{\matherr_j}[\mathX_j]}
\begin{equation}
\begin{aligned}
	\label{eq:MAP_optimization}
 \mathXMAP&= \mathargmax{\mathX}\prod_{j=1}^r \func{\exp}[-\frac{1}{2} \norm{\temp}[\mathinfo_j]]  
 \end{aligned}
\end{equation}
where $\exp{()}$  represents the exponential function, $\norm{e}[\mathinfo] = \mathtrs{e} \mathinfo {e}$ denotes the Mahalanobis norm, and $\mathtrs{.}$ is the transpose operator for vectors or matrices. The term $\matherr_j$  is an error function associated with the edge $j$, which depends on a subset of the variable nodes, and $\mathinfo_j$ is the information matrix of edge $j$, computed as the inverse of the covariance matrix.

Since maximizing the MAP objective is equivalent to minimizing the negative log-likelihood, this optimization problem can be reformulated as a weighted least squares problem \cite[\S III.B]{abdelkarim2025factor}
\begin{equation}
\label{eq:factor_graph_unconstrainted_optimization}
     \mathminimize{\mathX} \quad
        \sum_{j=1}^r   \norm{\temp}[\mathinfo_j], 
 \end{equation}
where  \mathX \; is the vector containing all variable nodes. In the following sections, we present solutions to this optimization problem, first without constraints and then with incorporated equality constraints.

\subsection{Gauss-Newton for  Unconstrained Least-Squares}
In factor graphs, edges typically define a nonlinear error function. Thus, optimization methods operate on a linearized version of this function. This is achieved through first-order Taylor expansion, given by
\deltemp
\begin{equation}
\begin{aligned}
 {\matherr_j}(\underbrace{\mathXinitial{\mathX_j}+ \Delta\mathX_j}_{\mathX_j}) 
 & \approx \func{\hat{\matherr}_j}[\Delta\mathX_j] = { \matherr_j(\mathXinitial{\mathX_j}) + \mathJacobian{e_j}{\mathXinitial{\mathX_j}} \Delta\mathX_j  }
\label{eq:linearlization}
\end{aligned}
\end{equation}
where $\mathXinitial{\mathX_j}$ is a linearization point and $\mathJacobian{e_j}{\mathXinitial{\mathX_j}}$ is the Jacobian matrix of the error function at edge $j$ evaluated at the linearization point  $\mathXinitial{\mathX_j}$. \\
Applying the stationarity condition (the gradient of the cost function vanishes at an optimal point), the Gauss-Newton update step at iteration $i$ is computed as  
\begin{equation}
\label{eq:sostepg2o}
 \underbrace{{\func{\sum_{j=1}^r} \text{map}(\text{H}_j^i})}_{\text{H}^i}\step[gn]^i =  \underbrace{\sum_{j=1}^r \text{map}(\mathfrak{b}^i_j)}_{\mathfrak{b}^i},  
\end{equation}
where the contribution of edge $j$ at iteration $i$ in building the linear system is 
\begin{subequations}
\begin{equation}
     \text{H}_j^i  =  \norm{  \mathJacobian{e_j}{{\mathX_j^i}}}[\mathinfo_j] 
\end{equation}
  \begin{equation}
     \mathfrak{b}^i_j =  - \mathtrs{ \mathJacobian{e_j}{{\mathX_j^i}}} \mathinfo_j \matherr_j ({\mathX}^i_j),
\end{equation}
\label{eq:cost_edge_contribution}
\end{subequations}
and map($\cdot$) is an operator that maps the local $\text{H}_j^i$ and $\mathfrak{b}^i_j$ to the global $\text{H}^i$ and $\mathfrak{b}^i$. This mapping accounts for the fact that error functions generally depend only on subsets of the variable nodes, and map($\cdot$) ensures proper zero-padding where necessary. 

Constructing and solving this linear system is a key computational step in factor graph optimization. Once the linear system is solved, the variable nodes are updated in each iteration as
\begin{equation}
\mathX^{i+1} = \mathX^i +  \Delta \mathX_{gn}^i
\end{equation}

Although the linear system formulation remains the same across unconstrained optimization methods, different techniques introduce modifications. For example, Levenberg-Marquardt improves numerical stability by adding a positive damping term to the diagonal elements of the matrix $\text{H}$, mitigating issues related to ill-conditioned matrices.

\section{Methodology}
We consider the optimization problem formulated in (\ref{eq:factor_graph_unconstrainted_optimization}), now extended to incorporate equality constraints, i.e.
\begin{subequations}
    \begin{alignat}{5}
        &\mathminimize{\mathX}     &\; \; & \sum_{j=1}^r\norm{\matherr_{j}(\mathX_j)}[\mathinfo_{j}]\\
        &\text{s.t.} \; &           &h_n(\mathX_n) =   0&   \; \; \qquad n  =1, \dots, l 
    \end{alignat}
\end{subequations}
where we assume the equality constraints to be linearly independent. The associated Lagrangian function \cite[\S \; 2.4.2]{Abdelkarim2020Development} is given by
\begin{equation}
    L(\mathX,\gamma)  =  \sum_{j=1}^r\norm{\matherr_{j}(\mathX_j)}[\mathinfo_{j}] +\sum_{n=1}^l\mathtrs{\gamma_{h_n}} h_n(\mathX_n),
\end{equation}
where the first term corresponds to the standard cost function in factor graphs, associated with the (regular) edges, while the second term introduces Lagrange multipliers $\gamma_{h_n}$ to enforce the equality constraints. 

This section provides a brief overview of the two methods implemented in \texttt{ecg2o}: the Augmented Lagrangian (AL) method and the Karush-Kuhn-Tucker (KKT)-based approach.

\subsection{Augmented Lagrangian Method}
The Augmented Lagrangian (AL) method enforces equality constraints by adding a penalty term to the Lagrangian function, resulting in the following formulation:
\begin{equation} 
 L_{\text{aug}}(\mathX,\gamma)  = L(\mathX,\gamma) +\sum_{n=1}^l\norm{h_n(\mathX_n)}[\text{P}_{h_n}]  		  
\end{equation}
where the penalty term is weighted by the matrix
\begin{equation}
    \text{P}_{h_n} = \text{diag}(\rho_{h_n,1}, \dots, \rho_{h_n,d}),
\end{equation}
where $\text{diag}(\cdot)$ represents a diagonal matrix of dimension $d$. The penalty parameter $\rho_.$ can either be uniform across all constraints or vary based on the algorithm's settings. Here, we assume a uniform $\rho_.$ for all constraints.

Notably, in AL, the penalty terms do not need to be excessively large, as discussed in \cite[Example 17.4]{wright2006numerical}. The algorithm iteratively minimizes the Augmented Lagrangian function with respect to  
$\mathX$, using the stationary condition, which yields an update step similar to (\ref{eq:sostepg2o}). 

While the contribution of regular edges remains unchanged (as described in (\ref{eq:cost_edge_contribution})), the contribution of equality edge $n$ at iteration $i$ can be expressed as:
\begin{subequations}
\begin{equation}
     \text{H}_n^i  =  \norm{  \mathJacobian{h_n}{{\mathX_n^i}} }[\text{P}_{h_n}]
\end{equation}
\begin{equation}
     \mathfrak{b}^i_n =  -\mathtrs{\mathJacobian{h_n}(\mathX^i_n)}\text{P}_{h_n} h_n(\mathX^i_n) - \mathtrs{\mathJacobian{h_n}(\mathX^i_n)} \gamma_{h_n}.
\end{equation}
\end{subequations}

The AL algorithm implemented in the \texttt{ecg2o} library iteratively solves the inner-loop unconstrained optimization while incorporating both cost and equality edge contributions when constructing the linear system. Upon satisfying the termination criteria or reaching the maximum number of inner-loop iterations, the algorithm exits the inner loop and updates the Lagrange multipliers
\begin{equation}
    	\gamma_{h_n}^{i+1} = 	\gamma_{h_n}^{i}  + \text{P}_{h_n} h_n(\mathX^i_n),
\end{equation}
along with the penalty parameters
\begin{equation}
    	\rho_.  = \max(\rho_{max}, 	\alpha 	\rho_.),
\end{equation}
where $\rho_{max}$ is the upper bound on the penalty parameter, and $\alpha \geq 1$ is the penalty update factor.

\subsection{KKT System-Based Gauss-Newton Method}
In this section, we present a method for defining a regular (cost) edge to represent equality constraints within a factor graph, leveraging the Karush-Kuhn-Tucker (KKT) system for step updates. 

Since both the error functions and equality constraints might be highly nonlinear, we consider their linearized forms using a first-order Taylor expansion. The Lagrangian function for the linearized system is given by:
\begin{equation}
  \hat{L}(\Delta\mathX,\gamma)  =  \sum_{j=1}^r\norm{\func{\hat{\matherr}_j}[\Delta\mathX_j] }[\mathinfo_{j}] +\sum_{n=1}^l\mathtrs{\gamma_{h_n}} \hat{h}_n(\Delta\mathX_n).
\end{equation}

Let $\Delta\mathX^*, \gamma^*$ be the local optimal points. These must satisfy the KKT conditions \cite[\S 5.3.3]{boyd2004convex} of the linearized optimization problem:
\begin{equation}
   \nabla_{\Delta \mathX}  \hat{L}(\Delta\mathX^*,\gamma^*) = 0, \quad \hat{h}(\Delta\mathX^*) = 0,
\end{equation}
where $\nabla$ denotes the gradient, $\hat{h}$ is the vector of all equality constraints. 

Assuming $\gamma^* = \gamma^i + \Delta\gamma^i$, the KKT system is formulated as:
\begin{equation}
\label{eq:KKT system for equality constraints}
\underbrace{\left\lbrack \begin{array}{cc}
	\text{H}^i& \textcolor{red}{\mathtrs{\mathJacobian{h}(\mathX^i)}} \\
	\textcolor{red}{\mathJacobian{h}(\mathX^i)} &\textcolor{red}{ 0 }
	\end{array}\right\rbrack }_{\text{KKT matrix}} 
\underbrace{\left\lbrack \begin{array}{c}
	\step[gn]^i  \\
	\textcolor{red}{\Delta \gamma^i} 
	\end{array}\right\rbrack }_{\text{update step}} 
= \underbrace{\left\lbrack \begin{array}{c}
		\mathfrak{b}^i \textcolor{red}{- \mathJacobian{h}^T (\mathX^i)\gamma^i} \\
	\textcolor{red}{-h(\mathX^i)}
	\end{array}\right\rbrack}_{\text{KKT vector}}
\end{equation}

The introduction of equality constraints modifies the linear system compared to unconstrained optimization, as highlighted in red. The key challenge lies in representing these constraints within the factor graph using appropriate variables and edges. Specifically, how can we define the error function and the weighting matrix associated with the equality constraints to enforce them effectively?
\\\\
\noindent\textbf{Equality Constraints as Factor Graph Edges}\\
We propose defining a regular edge for each equality constraint such that it produces the same update step as in (\ref{eq:KKT system for equality constraints}). Specifically, for an equality constraint $h_n$, we define a regular edge with the following error function and weighting matrix:
\begin{equation}
\label{eq:equality_error}
 \matherr_{h_n} = \left\lbrack \begin{array}{c}
		h_n(\mathX_n) \\\gamma_{h_n}
	\end{array}\right\rbrack, 
\quad
\mathinfo_{h_n} = \left\lbrack \begin{array}{cc} 
		 0_{d\times d} & \text{I}_{d\times d} \\
		 \text{I}_{d\times d}  & 0_{d\times d}
	\end{array}\right\rbrack,
\end{equation}
where $h_n$ and $\gamma_{h_n}$  have the dimension $d$, and $\text{I}$ denotes the identity matrix. This formulation results in a linear system equivalent to the KKT system of the equality-constrained optimization problem. Consequently, it allows us to enforce equality constraints while maintaining an unconstrained factor graph structure. Furthermore, in this approach, the Lagrange multipliers are treated as variable nodes, naturally integrating them into the optimization process.

For ease of implementation, the \texttt{ecg2o} library provides a dedicated class for equality edges, which automatically incorporates Lagrange multipliers and defines the associated weighting matrix. Additionally, the class simplifies the Jacobian implementation for equality edges, making the equality implementation as straightforward as for regular edges.

This design ensures an efficient and flexible extension of factor graphs, maintaining the computational advantages of existing optimization techniques while enabling constrained optimization in a natural manner.
\section{Results and Evaluation}
\label{sec:results_and_evaluation}
In this section, we present an optimal control problem that generates a sequence of control force inputs for an autonomous car to track a reference velocity trajectory. Our objective is to compare the performance of the Augmented Lagrange (AL) method with our proposed approach, which models equality constraints as regular edges by incorporating Lagrange multipliers.

\subsection{Problem Formulation}
Inspired by \cite{jia2023performance}, we formulate the optimal control problem as follows: 
\begin{subequations}
\label{eq:optimal_control}
    \begin{alignat}{5}
        &\mathminimize{}     &\;  & (\norm{x_N-r_N}[\text{P}]   + \sum_{k=1}^{N-1}(\norm{x_k-r_k}[\text{Q}] + \norm{u_k}[\text{R}])\\
        &\text{s.t.} \; &           & {x_{k+1} - \left[ x_k + \frac{\delta t}{m} (u_k - F_{\text{resis}}) \right]} = 0, \; k  =0, \dots, N-1 
    \end{alignat}
\end{subequations}
where $x_k$ represents the vehicle velocity, $u_k$ is the input force applied to the car (measured in Newtons), and 
$r_k$ is the reference velocity. In addition, the subscript  $k$ denotes the time instance, while $\delta t$ represents the sampling time of the controller, which defines the horizon length.

The resistance force $F_{\text{resis}}$ depends on the gravitational force $F_{\text{grav}}$, the rolling resistance $F_{\text{roll}}$, and the aerodynamic resistance force $F_{\text{air}}$. The total resistance force is given by:

\begin{equation}
  F_{resist} = \underbrace{m_vg\text{sin}\theta}_{F\text{grav}} +\underbrace{ \frac{1}{2} \rho_a A_f c_a x^2        }_{F\text{air}} + \underbrace{m_vgc_r\text{cos}\theta }_{F\text{roll}}
  \end{equation} 

The model is nonlinear due to the aerodynamic resistance force. The model parameters are as follows: $m$ represents the vehicle mass, including the effect of the rotational mass of the powertrain, while $m_v$ denotes the vehicle mass. The gravitational constant is given by  $g$, and  $\theta$ represents the road slope. The air density is denoted by  $\rho_a$, and $A_f$ is the frontal area of the vehicle. The coefficients   $c_a$  and $c_r$ correspond to the air drag and rolling resistance, respectively.

The factor graph representing the optimal control problem in (\ref{eq:optimal_control}) is illustrated in Fig. \ref{fig:oc_factor_graph}, where the equality constraints are incorporated using our proposed approach.

\begin{figure}[h]
    \centering
    \includegraphics{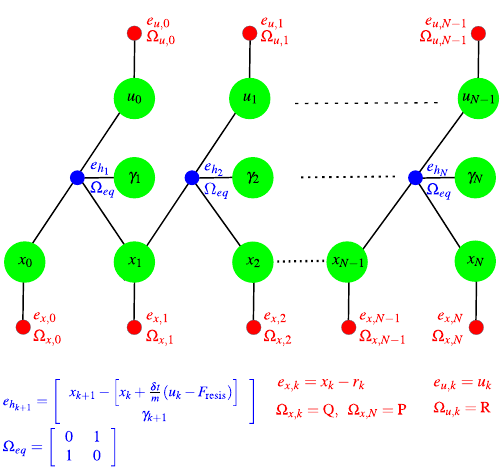}
    \caption{The factor graph for the optimal control problem. The red circles represent the cost terms as regular edges, while the blue circles illustrate the representation of equality constraints as regular edges. $k \in \{0, \cdots, N-1$\}}
    \label{fig:oc_factor_graph}
\end{figure}

\subsection{Evaluation}
  The reference trajectory for the velocity profile consists of 385 points with a sampling rate of 1 second. Using the factor graph-based optimization method, we solve the optimal control problem with weighting parameters $Q=P=1000$ and $ R=0.0007$  to determine the optimal input sequence. Applying this control sequence results in the optimal velocity profile. The velocity profile corresponding to the optimal control sequence obtained from solving the optimization problem is illustrated in Fig. \ref{fig:enter-label}. While different controller performances can be achieved with varying parameter settings, our primary focus is to evaluate the optimizer itself. The solution of the optimization problem using the Augmented Lagrangian method results in a velocity trajectory that is quite similar to our proposed approach. This is because both methods share the same stopping criteria.
  We computed the root mean squared error (RMSE) between both trajectories to quantify their similarity. The RMSE between the velocity trajectories obtained from the Augmented Lagrangian method and our proposed approach is 0.2426, indicating a high degree of similarity.
 
\begin{figure}[h]
    \centering
    \includegraphics[width=.9\linewidth]{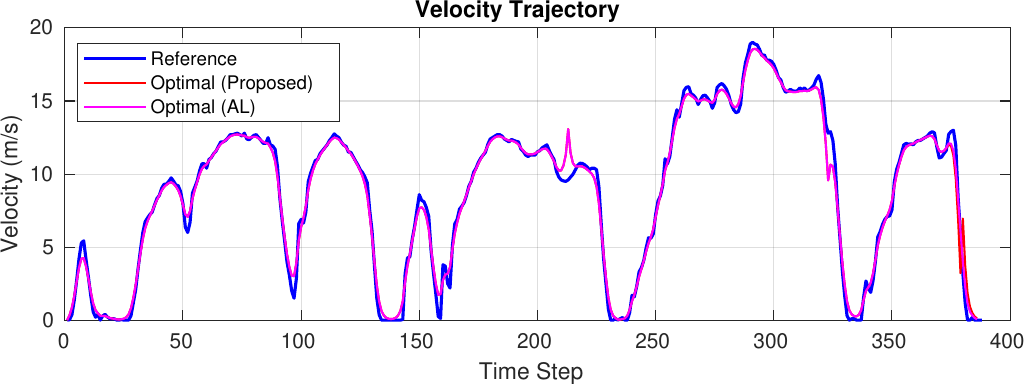}
    \caption{Optimal velocity trajectory obtained from the factor graph-based optimal control problem.}
    \label{fig:enter-label}
\end{figure}

 \noindent\textbf{Iteration Numbers}\\
In our case, the presence of nonlinear terms in the equality constraints might influence the number of iterations required for convergence. To address this effect, we considered an approximation for the nonlinear aerodynamic resistance term using a linear model:

\begin{equation} x^2 = p_1 + p_2 x, \end{equation}
where $p_1$ and $p_2$ are constants. 
Therefore, we considered five distinct scenarios for comparison.  First, we solve the unconstrained optimization problem, where the constraints are ignored, serving as a baseline to understand the effect of enforcing constraints. Next, we apply our proposed method to both a linearized dynamic model and the nonlinear system, allowing us to analyze how constraint approximation influences performance. Additionally, we compare these results with the Augmented Lagrangian (AL) approach, solving the problem in both linearized and nonlinear forms. In addition, we consider three different reference trajectory lengths—5, 100, and 385—to evaluate the optimization performance across varying problem sizes.
The results of the iteration number across all scenarios for the three trajectory lengths are summarized in Fig. \ref{fig:iteration_number}.
\begin{figure}[h]
    \centering
    \includegraphics[width=1
    \linewidth]{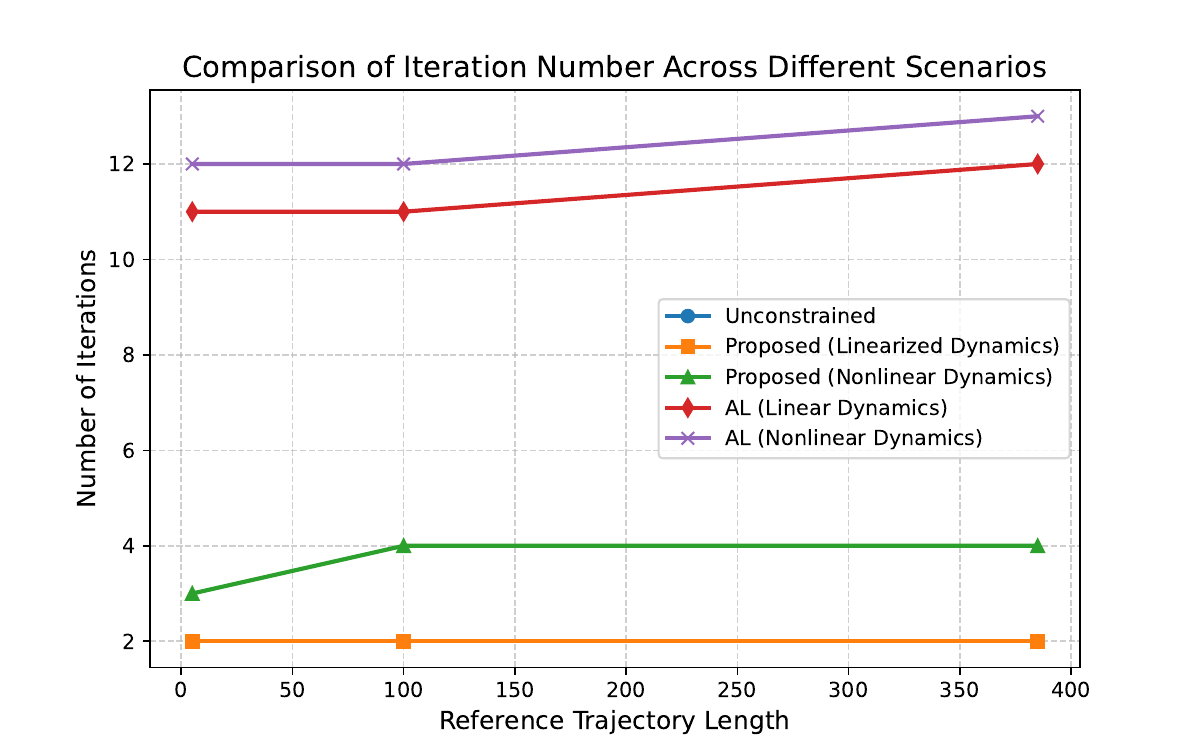}
    \caption{Iteration number across different optimization scenarios and trajectory lengths. }
    \label{fig:iteration_number}
\end{figure}

The Fig \ref{fig:iteration_number} shows that the iteration number in the unconstrained optimization matches that of our proposed method with linearized dynamics, converging in just two iterations across all trajectory lengths. For nonlinear dynamics, our method requires slightly more iterations (3–4), reflecting the added complexity while maintaining efficiency.

In contrast, the Augmented Lagrangian (AL) method requires 11–13 iterations, approximately three times more than our proposed method, highlighting its higher computational cost. These results were obtained after careful parameter tuning, where we set the maximum number of inner iterations to 1, with $\rho_{\text{init}}=10$, $\rho_{\text{max}}=50000$, and $\alpha=10$. However, further tuning is still required for different optimization problems.

Regarding sensitivity to the initial value of the Lagrange multiplier, we found that both algorithms maintained a stable iteration number across different Lagrange multiplier initial values.  

\noindent\textbf{Computation time}\\
We compare the computation time for the nonlinear dynamics scenario across three different trajectory lengths. The optimization problem is solved 1,000 times, and the reported computation time represents the average over these runs to ensure consistency. All experiments were conducted on a Linux system with an Intel Core i9 12th Gen processor and 32 GB of RAM. The results are shown in Fig. \ref{fig:computation_time}

\begin{figure}[h]
    \centering
    \includegraphics[width=.95\linewidth]{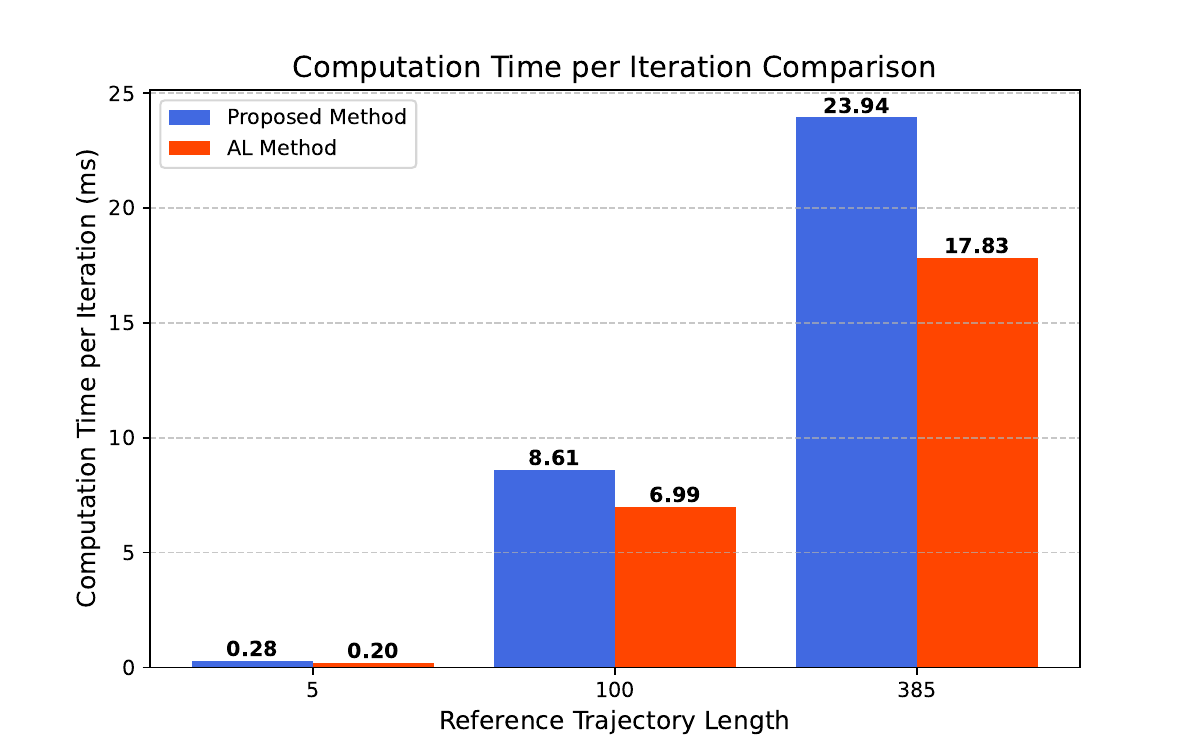}
    \caption{Computation time per iteration for different trajectory lengths using the Proposed Method and Augmented Lagrangian (AL) Method. }
    \label{fig:computation_time}
\end{figure}
The computation time for the AL method is slightly lower than that of our proposed method, which was expected since our solver handles a larger linear system in each iteration. However, advancements in linear solvers have reduced the impact of solving larger systems on overall computation time, making the difference less apparent. The increase in computation time for our method ranges between $20\%$ and $35\%$, but the significant reduction in iteration count makes it a worthwhile trade-off, especially for applications requiring fast convergence.

\section{Conclusion and Future Work}
\label{sec:conclusion}
 In this paper, we addressed the challenge of incorporating equality constraints into factor graphs, an essential extension for applications beyond robotic perception, such as optimal control. While the Augmented Lagrangian (AL) method is the state-of-the-art approach for handling equality constraints, we proposed an alternative method that natively integrates constraints within factor graph optimization without requiring additional iterative adjustments.

Our proposed approach outperforms AL by achieving faster convergence and eliminating the need for parameter tuning, making it more practical. We validated our method through a case study on velocity tracking in autonomous vehicles, demonstrating its efficiency and robustness.

As future work, we aim to explore more advanced constrained optimization techniques beyond AL, specifically targeting the handling of inequality constraints within factor graph frameworks. This would further expand the applicability of factor graphs to  control and planning problems in robotics.

\addtolength{\textheight}{-12cm}   % This command serves to balance the column lengths

%%%%%%%%%%%%%%%%%%%%%%%%%%%%%%%%%%%%%%%%%%%%%%%%%%%%%%%%%%%%%%%%%%%%%%%%%%%%%%%%

\bibliographystyle{IEEEtran} 
\bibliography{section/10-bibliography.bib}

\begin{thebibliography}{10}
\providecommand{\url}[1]{#1}
\csname url@samestyle\endcsname
\providecommand{\newblock}{\relax}
\providecommand{\bibinfo}[2]{#2}
\providecommand{\BIBentrySTDinterwordspacing}{\spaceskip=0pt\relax}
\providecommand{\BIBentryALTinterwordstretchfactor}{4}
\providecommand{\BIBentryALTinterwordspacing}{\spaceskip=\fontdimen2\font plus
\BIBentryALTinterwordstretchfactor\fontdimen3\font minus
  \fontdimen4\font\relax}
\providecommand{\BIBforeignlanguage}[2]{{%
\expandafter\ifx\csname l@#1\endcsname\relax
\typeout{** WARNING: IEEEtran.bst: No hyphenation pattern has been}%
\typeout{** loaded for the language `#1'. Using the pattern for}%
\typeout{** the default language instead.}%
\else
\language=\csname l@#1\endcsname
\fi
#2}}
\providecommand{\BIBdecl}{\relax}
\BIBdecl

\bibitem{dellaert2017factor}
F.~Dellaert and M.~Kaess, ``Factor graphs for robot perception,''
  \emph{Foundations and Trends{\textregistered} in Robotics}, vol.~6, no. 1-2,
  pp. 1--139, 2017.

\bibitem{tourani2022visual}
A.~Tourani, H.~Bavle, J.~L. Sanchez-Lopez, and H.~Voos, ``Visual slam: What are
  the current trends and what to expect?'' \emph{Sensors}, vol.~22, no.~23, p.
  9297, 2022.

\bibitem{bavle2022s}
H.~Bavle, J.~L. Sanchez-Lopez, M.~Shaheer, J.~Civera, and H.~Voos, ``S-graphs+:
  Real-time localization and mapping leveraging hierarchical representations,''
  \emph{IEEE Robotics and Automation Letters}, 2022.

\bibitem{lai2017solving}
W.~H. Lai, S.~L. Kek, and K.~G. Tay, ``Solving nonlinear least squares problem
  using gauss-newton method,'' \emph{International Journal of Innovative
  Science, Engineering \& Technology}, vol.~4, no.~1, pp. 258--262, 2017.

\bibitem{more2006levenberg}
J.~J. Mor{\'e}, ``The levenberg-marquardt algorithm: implementation and
  theory,'' in \emph{Numerical analysis: proceedings of the biennial Conference
  held at Dundee, June 28--July 1, 1977}.\hskip 1em plus 0.5em minus
  0.4em\relax Springer, 2006, pp. 105--116.

\bibitem{gtsam_}
F.~Dellaert and G.~Contributors, ``borglab/gtsam,'' 2022, available online at:
  \url{https://github.com/borglab/gtsam}, accessed: 2024-06-13.

\bibitem{kummerle2011g}
R.~K{\"u}mmerle, G.~Grisetti, H.~Strasdat, K.~Konolige, and W.~Burgard, ``g2o:
  A general framework for graph optimization,'' in \emph{2011 IEEE
  International Conference on Robotics and Automation}.\hskip 1em plus 0.5em
  minus 0.4em\relax IEEE, 2011, pp. 3607--3613.

\bibitem{grisetti2020least}
G.~Grisetti, T.~Guadagnino, I.~Aloise, M.~Colosi, B.~Della~Corte, and
  D.~Schlegel, ``Least squares optimization: from theory to practice,''
  \emph{Robotics}, vol.~9, no.~3, p.~51, July 2020.

\bibitem{abdelkarim2025factor}
A.~Abdelkarim, H.~Voos, and D.~G{\"o}rges, ``Factor graphs in
  optimization-based robotic control-a tutorial and review,'' \emph{IEEE
  Access}, 2025.

\bibitem{andersson2019casadi}
J.~A. Andersson, J.~Gillis, G.~Horn, J.~B. Rawlings, and M.~Diehl, ``Casadi: a
  software framework for nonlinear optimization and optimal control,''
  \emph{Mathematical Programming Computation}, vol.~11, pp. 1--36, 2019.

\bibitem{abdelkarim2020optimal}
A.~Abdelkarim and P.~Zhang, ``Optimal scheduling of preventive maintenance for
  safety instrumented systems based on mixed-integer programming,'' in
  \emph{Model-Based Safety and Assessment: 7th International Symposium, IMBSA
  2020, Lisbon, Portugal, September 14--16, 2020, Proceedings 7}.\hskip 1em
  plus 0.5em minus 0.4em\relax Springer, 2020, pp. 83--96.

\bibitem{abdelkarim2023optimization}
A.~Abdelkarim, Y.~Jia, and D.~Gorges, ``Optimization of vehicle-to-grid
  profiles for peak shaving in microgrids considering battery health,'' in
  \emph{IECON 2023-49th Annual Conference of the IEEE Industrial Electronics
  Society}.\hskip 1em plus 0.5em minus 0.4em\relax IEEE, 2023, pp. 1--6.

\bibitem{wachter2006implementation}
A.~W{\"a}chter and L.~T. Biegler, ``On the implementation of an interior-point
  filter line-search algorithm for large-scale nonlinear programming,''
  \emph{Mathematical programming}, vol. 106, pp. 25--57, 2006.

\bibitem{chen2019LQR}
G.~Chen and F.~Zhang, Yetongand~Dellaert, ``Lqr control using factor graphs,''
  2019, available online at:
  \url{https://gtsam.org/2019/11/07/lqr-control.html},accessed: 2024-06-15.

\bibitem{yang2021equality}
S.~Yang, G.~Chen, Y.~Zhang, H.~Choset, and F.~Dellaert, ``Equality constrained
  linear optimal control with factor graphs,'' in \emph{2021 IEEE International
  Conference on Robotics and Automation (ICRA)}.\hskip 1em plus 0.5em minus
  0.4em\relax IEEE, 2021, pp. 9717--9723.

\bibitem{darnley2021flow}
R.~Darnley, ``Flow control of wireless mesh networks using lqr and factor
  graphs,'' master thesis, Carnegie Mellon University, 2021.

\bibitem{kim2023simultaneous}
S.~Kim, D.~K. Jha, D.~Romeres, P.~Patre, and A.~Rodriguez, ``Simultaneous
  tactile estimation and control of extrinsic contact,'' in \emph{2023 IEEE
  International Conference on Robotics and Automation (ICRA)}.\hskip 1em plus
  0.5em minus 0.4em\relax IEEE, 2023, pp. 12\,563--12\,569.

\bibitem{chen2022gtgraffiti}
G.~Chen, S.~Baek, J.-D. Florez, W.~Qian, S.-w. Leigh, S.~Hutchinson, and
  F.~Dellaert, ``Gtgraffiti: Spray painting graffiti art from human painting
  motions with a cable driven parallel robot,'' in \emph{2022 International
  Conference on Robotics and Automation (ICRA)}.\hskip 1em plus 0.5em minus
  0.4em\relax IEEE, 2022, pp. 4065--4072.

\bibitem{chen2022locally}
G.~Chen, S.~Hutchinson, and F.~Dellaert, ``Locally optimal estimation and
  control of cable driven parallel robots using time varying linear quadratic
  gaussian control,'' in \emph{2022 IEEE/RSJ International Conference on
  Intelligent Robots and Systems (IROS)}.\hskip 1em plus 0.5em minus
  0.4em\relax IEEE, 2022, pp. 7367--7374.

\bibitem{bertsekas2014constrained}
D.~P. Bertsekas, \emph{Constrained optimization and Lagrange multiplier
  methods}.\hskip 1em plus 0.5em minus 0.4em\relax Academic press, 2014.

\bibitem{sodhi2020ics}
P.~Sodhi, S.~Choudhury, J.~G. Mangelson, and M.~Kaess, ``Ics: Incremental
  constrained smoothing for state estimation,'' in \emph{2020 IEEE
  International Conference on Robotics and Automation (ICRA)}.\hskip 1em plus
  0.5em minus 0.4em\relax IEEE, 2020, pp. 279--285.

\bibitem{qadri2022incopt}
M.~Qadri, P.~Sodhi, J.~G. Mangelson, F.~Dellaert, and M.~Kaess, ``Incopt:
  Incremental constrained optimization using the bayes tree,'' in \emph{2022
  IEEE/RSJ International Conference on Intelligent Robots and Systems
  (IROS)}.\hskip 1em plus 0.5em minus 0.4em\relax IEEE, 2022, pp. 6381--6388.

\bibitem{bazzana2022handling}
B.~Bazzana, T.~Guadagnino, and G.~Grisetti, ``Handling constrained optimization
  in factor graphs for autonomous navigation,'' \emph{IEEE Robotics and
  Automation Letters}, vol.~8, no.~1, pp. 432--439, 2022.

\bibitem{bazzana2024augmented}
B.~Bazzana, H.~Andreasson, and G.~Grisetti, ``How-to augmented lagrangian on
  factor graphs,'' \emph{IEEE Robotics and Automation Letters}, 2024.

\bibitem{Abdelkarim2020Development}
A.~Abdelkarim, ``Development of numerical solvers for online optimization with
  application to mpc-based energy-optimal adaptive cruise control,'' master
  thesis, Technische Universit{\"a}t Kaiserslautern, 2020. Available online at
  \url{http://dx.doi.org/10.13140/RG.2.2.11897.28000}, accessed: 2025-02-28.

\bibitem{wright2006numerical}
J.~Nocedal and S.~J. Wright, ``Numerical optimization,'' 2006.

\bibitem{boyd2004convex}
S.~P. Boyd and L.~Vandenberghe, \emph{Convex optimization}.\hskip 1em plus
  0.5em minus 0.4em\relax Cambridge university press, 2004.

\bibitem{jia2023performance}
Y.~Jia, A.~Abdelkarim, X.~Klingbeil, R.~Savelsberg, and D.~G{\"o}rges,
  ``Performance evaluation of energy-optimal adaptive cruise control in
  simulation and on a test track,'' \emph{IFAC-PapersOnLine}, vol.~56, no.~2,
  pp. 4994--5000, 2023.

\end{thebibliography}

\end{document}